\documentclass[
reprint,
superscriptaddress,
nofootinbib,
amsmath,
amssymb,
aps,
prl,
]{revtex4-1}

\usepackage[utf8]{inputenc}
\usepackage{amsmath}
\usepackage{mathtools}
\usepackage{amssymb,amstext,amsfonts}
\usepackage{hyperref}
\usepackage{graphicx}
\usepackage{float}
\usepackage{microtype}
\usepackage{subfigure}
\usepackage{multirow}
\usepackage{lineno}
\usepackage{color}
\usepackage[usenames,dvipsnames]{xcolor}
\usepackage{comment}
\usepackage[normalem]{ulem}

\newcommand{\be}{ \begin{equation} }
\newcommand{\ee}{ \end{equation}}

\begin{document}

\title{Bekenstein-Hod Universal Bound on Information Emission Rate Is Obeyed by LIGO-Virgo Binary Black Hole Remnants}

\author{Gregorio Carullo}
\affiliation{Dipartimento di Fisica ``Enrico Fermi'', Universit\`a di Pisa, Pisa I-56127, Italy}
\affiliation{INFN sezione di Pisa, Pisa I-56127, Italy}
\author{Danny Laghi}
\affiliation{Dipartimento di Fisica ``Enrico Fermi'', Universit\`a di Pisa, Pisa I-56127, Italy}
\affiliation{INFN sezione di Pisa, Pisa I-56127, Italy}
\author{John Veitch}
\affiliation{Institute for Gravitational Research, University of Glasgow, Glasgow, G12 8QQ, United Kingdom}
\author{Walter Del Pozzo}
\affiliation{Dipartimento di Fisica ``Enrico Fermi'', Universit\`a di Pisa, Pisa I-56127, Italy}
\affiliation{INFN sezione di Pisa, Pisa I-56127, Italy}

\date{\today}

\begin{abstract}
Causality and the generalized laws of black hole thermodynamics imply a bound, known as the \textit{Bekenstein--Hod universal bound}, on the information emission rate of a perturbed system. Using a time-domain ringdown analysis, we investigate whether remnant black holes produced by the coalescences observed by Advanced LIGO and Advanced Virgo obey this bound. We find that the bound is verified by the astrophysical black hole population with $94\%$ probability, providing a first confirmation of the Bekenstein--Hod bound from black hole systems.
\end{abstract}

\maketitle

\textit{Introduction--} Black holes (BHs) are fascinating objects 
that have been a fruitful ground for theoretical exploration of the frontiers of physics. 
Their consideration within the general theory of relativity (GR) has led to many surprising results 
including the inescapable presence of a singularity inside the horizon, 
the cosmic censorship conjecture, BH uniqueness and final state conjecture, 
break--down of classical physics near a singularity, and the information paradox~\cite{Penrose1965,Hawking,Ginzburg,Zeldovich,Israel,Carter,Hawking1972,Robinson,Mazur,Bunting, Pretorius2017, Dafermos:2008en, Misner1973, Hawking1975}. 
They are thus of central interest for hints of ``new physics'';
BH dynamical processes could hint to new particles~\cite{Brito:2014wla} or to deviations from semiclassical 
physics predictions already at macroscopic distances from the horizon~\cite{Almheiri:2012rt, Hod:2011aa}.
The advent of gravitational--wave (GW) astronomy has given access to regimes
where the unique properties of BHs can be observationally tested.
Here, we present a new observational result on the information emission mechanism of perturbed BHs.

The information emission rate of a physical system depends on its geometry, on the message carrier,
and on the dynamics responsible for the signal. 
Despite the details, a universal limit on the information rate exists~\cite{Hod:2006jw}. 
Hod's argument stems from a result due to Bekenstein~\cite{Bekenstein:1981zz}
concerning the maximum amount of information that can be stored in a static system.
Assuming the generalized second law of thermodynamics and
the principle of causality, Bekenstein derived a universal bound for the maximum 
average \emph{rate of information emission} achievable by a physical system:
\begin{equation}\label{eq:Bek}
\dot{I}_{max} = \frac{\pi \, E}{\hbar \, \ln 2} \,,
\end{equation}
where $E$ is the signal energy in the receiver's frame and the dot stands for time derivative.
A similar bound, whose validity was restricted to noisy channels, was obtained by Bremermann \cite{Bremermann}, 
building upon the foundational work of Shannon \cite{ShannonWeaver49}.
Hod argued that Eq.~(\ref{eq:Bek}) could be recast in the form~\cite{Hod:2006jw}:
\begin{equation}\label{eq:Hod}
\mathcal{H} := \frac{1}{ \pi} \cdot \frac{\hbar \, \omega_I}{k_B T} \leq 1\,,
\end{equation}
where $T$ is the temperature of the system under consideration and $\omega_I := 1/\tau$ is its inverse relaxation time.
This is the \textit{Bekenstein-Hod universal bound}.
Hod further noted that this result also applies to the ringdown phase of a non-extremal BH relaxing towards equilibrium.
As a consequence of the final state conjecture~\cite{Ginzburg,Zeldovich,Israel,Carter,Hawking1972,Robinson,Mazur,Bunting, Pretorius2017, Dafermos:2008en}, 
asymptotically-flat perturbed BHs radiate away all their anisotropies~\cite{Swetha,Okounkova2020}
and quickly relax towards a stationary Kerr-Newman spacetime, characterized by only three parameters: mass, spin and charge.
The corresponding GW signal, after an initial burst of radiation dependent on the perturbation,
can be represented as a sum of damped sinusoids forming a spectrum of discrete complex frequencies characteristic of the system. 
This spectrum is \textit{solely} determined by the intrinsic properties of the spacetime and does not depend on the details of the perturbation.
The corresponding spacetime modes are known as quasinormal modes (QNMs) and the emission process is referred to as \textit{ringdown}~\cite{Nollert_QNM, Ferrari_QNM, Kokkotas_QNM, QNM_review_BCS, Maggiore:2018sht}.
In the presence of a BH ringdown, the temperature $T$ in Eq.~\ref{eq:Hod} is the Hawking temperature \cite{Hawking1975} and the relaxation time in Eq.~(\ref{eq:Hod}) can be taken as the imaginary part of the least damped QNM \cite{Hod:2006jw}.
Assuming that the BH under consideration generates a Kerr spacetime, thus neglecting electrical charge,
the temperature can be expressed as a function of the BH mass and spin.

The direct observation of a ringdown phase provides an agnostic measurement of the relaxation time of the newly-born BH, and thus allows us to test Eq.~(\ref{eq:Hod}).
We proceed to perform a combined analysis of the remnant objects produced by the BBH mergers detected by the LIGO-Virgo Collaboration~\cite{AdvLIGO,AdvVirgo, GWTC-1, GWTC-2} to place the first experimental constraint on the Bekenstein-Hod bound.
The result we find is in agreement with the bound predicted by GR, BH thermodynamics, and information theory.\\Throughout this Letter, $c=G=1$ units are used. Unless otherwise specified, when discussing parameters point estimates, we quote median and $90\%$ credible levels.\\ \\

\textit{The bound in the literature-} Eq.~(\ref{eq:Hod}) is respected by classical linear perturbations on non-extremal BH spacetimes in GR, 
as shown by Hod~\cite{Hod:2006jw, Hod:2007tb,Hod:2008se,Hod:2009td,Hod:2018ifz}. 
Not only do BHs comply with this bound, they reach it in the limit of extremal spin, 
implying that BHs can be regarded as the fastest dissipating objects (for a given temperature) in the Universe~\cite{Hod:2007tb}.
In the extremal spin limit, the temperature of the BH approaches zero and the relaxation timescale increases in an unbounded fashion.
The colder a system gets, the longer it takes to return to its equilibrium state. This is a direct consequence of the third law of thermodynamics:
no system can reach a zero-temperature state in a finite amount of time. 
The equivalent law in BH dynamics, preventing a BH to reach its extremal spin value 
(which would also violate the Cosmic Censorship, see~\cite{hartle:2003} Chapter 15), 
was proven by Carter~\cite{PhysRevLett.57.397}.
The family of BHs that attains the fastest relaxation rate are
Reissner-Nordstrom BHs with $Q\simeq0.726$~\cite{Hod:2018ifz}.
A looser bound for classical systems (with the same functional form as the Bekenstein-Hod bound) 
can be derived through statistical arguments, the laws of quantum mechanics, and thermodynamics, see Ref.~\cite{Ropotenko:2007jm}.
Classical systems are not expected to saturate this bound, while systems where quantum fluctuations dominate do.
Interestingly, Kerr BHs very close to extremality can reach the bound in the Hod formulation, 
according to the results of Ref.~\cite{Yang:2013uba}.
Nevertheless, astrophysical BHs never attain such extreme values of the spin, and thus such behaviour is not relevant to observational analyses.
These considerations pose the question of whether thermal fluctuations of a BH can be treated thermodynamically 
or if they should be described similarly to quantum fluctuations~\cite{Pesci:2009xh}. 
They can indeed be treated thermodynamically~\cite{Ropotenko:2008ca}.
Although the ringdown of a perturbed BH is classical,
the bound implicitly assumes a quantum theory via Hawking's temperature~\cite{Hawking:1974rv} 
and the interpretation of the information as entropy, hence as a statistical quantity (e. g. Ref.~\cite{Jaynes2003}, Chapter 11).
A similar bound on the rate of chaos growth was derived in Ref.~\cite{Maldacena2016} 
within the context of thermal quantum systems.
The connection between this latter result and the Bekenstein-Hod bound is shown by
recalling that for test fields on an unstable circular null geodesics of a wide class of static, spherically symmetric, asymptotically flat spacetimes, in the eikonal limit the Lyapunov exponent of the system approaches the imaginary part of the QNM, that is, $\lambda \sim \omega_I$~\cite{Cardoso:2008bp} (for a general analysis on the validity of this connection and a counter-example in alternative theories of gravity, see Refs.\cite{Konoplya:2017wot, Konoplya:2019hml}). Finally, the bound has recently been shown to provide a proof of the Strong Cosmic Censorship Conjecture of dynamically formed BHs \cite{Hod:2020ktb}.\\
For discussions on the interpretations of the bound, see Ref.~\cite{Bekenstein:2004sh}.\\ \\

\textit{Analysis--}
The \textit{Bekenstein-Hod universal bound} has drawn much attention not only due to its connections between information theory and BH thermodynamics, 
but also in relation to the viscosity-entropy bound~\cite{Hod:2018fbj, Pesci:2008zv, Pesci:2009dc}.
However, an experimental verification of this bound on BH systems is still missing.
The discovery of GW from binary black holes (BBH) has given direct access to the intrinsic properties of dynamical spacetimes~\cite{GW150914, TGR_paper_GW150914, PE_paper_GW150914, Abbott:2016izl, GWTC-1, GWTC-1_TGR, GWTC-2, GWTC-2_TGR}.
GW observations constrain both the relaxation time (in an agnostic, semi-model independent fashion)
and the mass and spin characterizing the remnant Kerr spacetime, 
allowing to directly verify the information emission bound.
We will first detail our method using GW150914, 
and later extend our analysis to the BBHs reported from the LIGO-Virgo collaborations~\cite{GWTC-2_TGR}. \\

\textit{Analysis of GW150914--} 
The first BBH coalescence~\cite{GW150914} signal can be used to assess how an individual detection from current 
ground-based detectors constrains the Bekenstein-Hod bound. 

To verify Eq.~(\ref{eq:Hod}), the first ingredient is an agnostic estimation of the remnant relaxation time. 
This timescale drives the evolution of spacetime after the two progenitors coalesced and merged to form a single object,
which then approaches equilibrium through the emission of damped normal modes characteristic of the system.
We can obtain such an agnostic measurement employing the \texttt{pyRing} pipeline~\cite{PhysRevD.99.123029, Isi:2019aib, GWTC-2_TGR}. 
We describe LIGO data, $D$, around GW150914 as a zero-mean wide-sense-stationary gaussian 
process (see~\cite{TGR_paper_GW150914} for a validation of this hypothesis) and use a template described by a single damped sinusoid. 
The template choice reflects our semi-model-independent approach, 
i.e., we assume that the relaxation is driven by an exponential decay,
without imposing any other prediction of GR. We thus allow for the possibility that the remnant 
is a compact object mimicking a BH~\cite{Barack:2018yly,Raposo:2018rjn,Mazur:2004fk,Visser:2003ge} 
with properties, like the relaxation time and the ringdown emission amplitudes, that might differ from a Kerr BH.
Since the relaxation time of the BH is dictated by the longest lived mode, 
we ignore the full post-merger signal where shorter lived modes might play a role \cite{Isi:2019aib, Giesler:2019uxc, GWTC-2_TGR, Bhagwat:2019dtm, Forteza:2020hbw, Ota:2019bzl, CalderonBustillo:2020tjf, Okounkova:2020vwu, Loutrel:2020wbw, Mourier:2020mwa} 
and we begin the analysis at $t=10\,M_f$ after the GW strain peak, where $M_f$ is the mass of the remnant BH in geometrical units. 
This analysis yields $\tau = 4.8^{+3.7}_{-1.9} \, \, \mathrm{ms}$.
Full details can be found in Ref.~\cite{GWTC-2_TGR} and its associated data release~\cite{GWTC2:TGR:release}.

Next, we need an estimate of the remnant mass and spin.
Measurements of the initial masses and spins of the progenitors can be used to predict the remnant mass and adimensional spin ($a_f$), 
by modeling the emission of energy and angular momentum~\cite{UIB2016_FF_paper, surfinbh1, surfinbh2}.
We use public samples on progenitors parameters~\cite{O1_samples_release}, yielding $M_f = 68.0^{\, +3.8}_{\, -3.6} M_{\odot}$, $a_f = 0.68^{\, +0.05}_{\, -0.06}$.
Since the mass dependence simplifies between the damping time and the temperature, the bound does not directly depend on the redshift.
To avoid specifying a cosmology, all the quantities are consistently computed in the earth-based "detector frame", redshifted by a factor $(1+z)$ with respect to the frame of the emitting source.

\begin{figure}[t]
\includegraphics[width=0.5\textwidth]{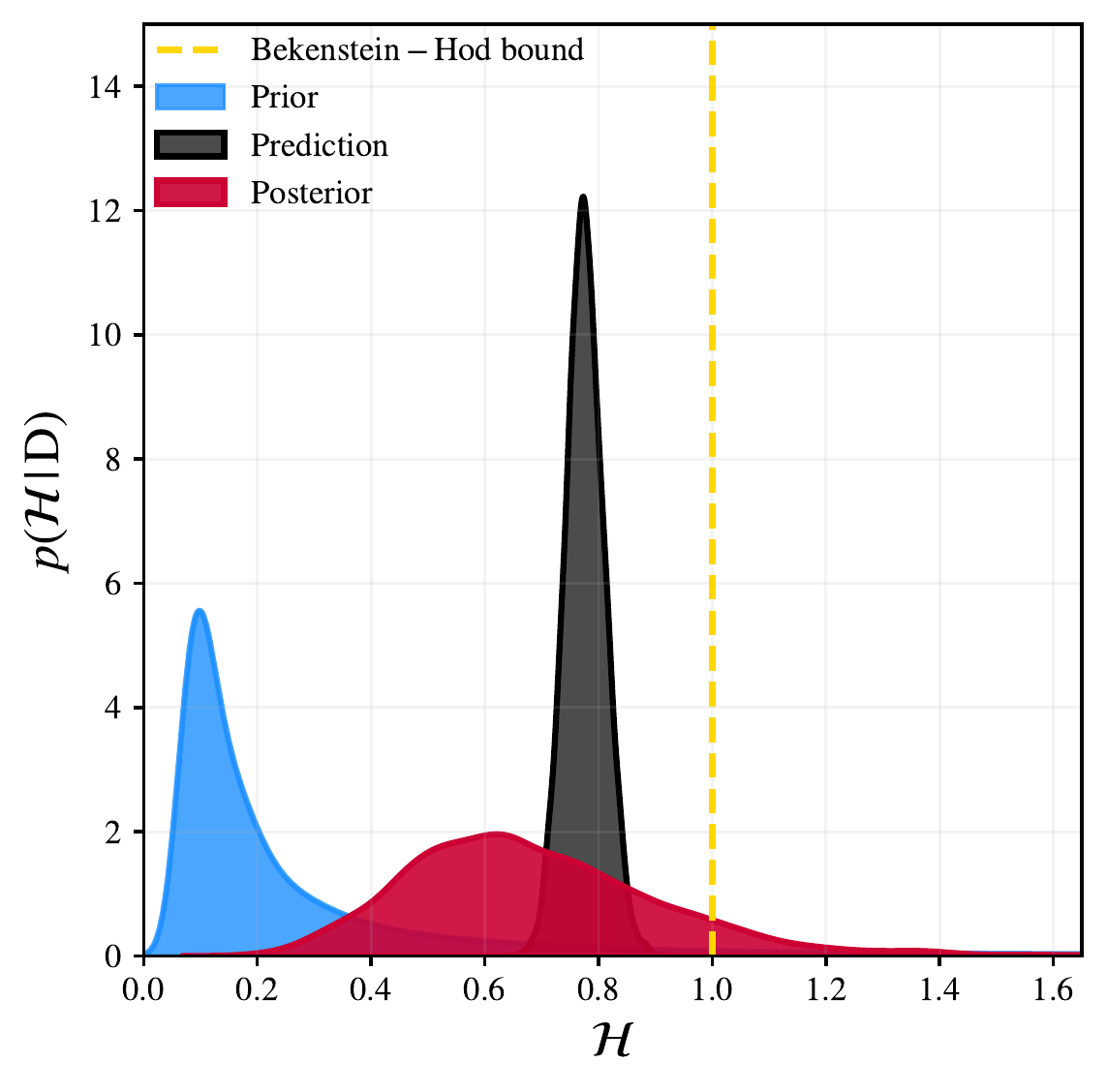}
\caption{Posterior probability distribution (red area) of the Bekenstein-Hod bound (gold vertical line). For comparison, the $\mathcal{H}$ distribution when enforcing GR predictions (black area) and the prior distribution (light blue area) are also displayed. The bound is satisfied at $91\%$ probability.}
\label{fig:bound} 
\end{figure}

From these values, we infer a posterior distribution for the $\mathcal{H}$ parameter (Fig.~(\ref{fig:bound}), red solid line), evaluating the probability that it obeys the constraint (Fig.~(\ref{fig:bound}), yellow vertical line). The probability $P(\mathcal{H}<1\, |\, \mathrm{D})$ is equal to $91\%$.
The statistical uncertainty is dominated by the measurement of the relaxation time.
This can be seen in the following way. Instead of measuring $\tau$, we can assume that the ringdown emission 
corresponds to the least-damped mode, $(\ell, |m|, n)=(2,2,0)$, as predicted by GR ~\cite{Vishveshwara, Chandrasekhar:1975zza, LISA_spectroscopy, QNM_review_BCS, Berti_website}, thus \textit{fixing} $\tau$ as a function of the remnant mass and spin, and eliminating its related uncertainty. This yields the black distribution shown in Fig.~(\ref{fig:bound}). 
As expected when enforcing GR relations, the values of $\mathcal{H}$ completely fall below the predicted bound, yielding a narrower distribution, confirming that the uncertainty is indeed dominated by the measurement of $\tau$. 
For comparison, we plot in blue the prior distribution imposed on $\mathcal{H}$ by the standard priors on the remnant parameters employed in the LIGO-Virgo collaboration (LVC) BBH analyses~\cite{GWTC-2,GWTC-2_TGR}. These are uniform on the damping time, while on the remnant mass and spin are determined by assuming a uniform prior on redshifted component masses and spin magnitudes and isotropic in spin orientations.
The prior distribution peaks at values consistent with the Bekenstein-Hod bound, but extends up to $\mathcal{H}=18$ with $9\%$ support above the bound. As expected for a loud event such as GW150914, the posterior distribution differs significantly from the prior. We discuss alternative prior choices in the next section.
Although the data from which the temperature was estimated has some overlap with the ringdown data, our conclusions are not affected. First of all, the hypotheses employed in estimating the temperature and the damping time are independent. Moreover, since the fraction of SNR contained in the late time ringdown (starting $10 \, M_f$ past the strain peak) is modest, the temperature estimate is largely independent of the last portion of the data.\\
Note that in \textit{ordinary} classical thermodynamical systems (i.e. not strongly self-gravitating compact systems), $\mathcal{H}$ would attain a value of the order $\sim10^{-11}$~\cite{Hod:2006jw}. The large values attained by the posterior distributions shown in Fig.~(\ref{fig:bound}) are an astonishing consequence of the peculiar characteristics of BH vacuum solutions.

\textit{Joint GWTC analysis --} Having verified the bound on GW150914, 
we proceed to investigate whether other BBH detections from LIGO and Virgo also respect the bound.
Not all of the events present in the GWTC-2 catalog \cite{GWTC-1, GWTC-2} show a detectable ringdown,
since low-mass systems merge at frequencies $O(k\text{Hz})$, where the sensitivity of the detector quickly degrades.
Hence, we repeat the ringdown analysis on the BBH events included in Ref.~\cite{GWTC-2_TGR}, applying two additional requirements: i) the posteriors of the $\mathcal{H}$ parameters differ from the assumed priors and ii) the logarithm of the Bayes factor between the hypothesis of a damped sinusoidal signal being present in the data versus the hypothesis that the data contain only noise, is positive ($\log B^{s}_{n} > 0$). These additional requirements are needed, since the selection criterion applied in Ref.~\cite{GWTC-2_TGR} was based on a different waveform template than the one adopted in our current analysis\footnote{Our analysis focuses on the \textit{observed} BH population properties, hence the treatment of selection effects is not relevant to our study.}.
We compute posterior distributions on the bound with the same procedure outlined for GW150914, using posterior samples presented in Refs.~\cite{GWTC-2, GWTC-2_TGR} and available at ~\cite{GWTC2:release, GWTC2:TGR:release}.
In Table~\ref{tab:probs}, we report the posterior probability $P(\mathcal{H}_i<1\, |\, \mathrm{D}_i)$ for each of the events using the method described in the previous section. 
The extension of the analysis to this pool of events, which now includes low signal-to-noise ratio signals, implies that our estimate will be more sensitive to the specific choice of prior employed in the analysis. To address this issue, in Table~\ref{tab:probs} we also list the probabilities extracted from single events likelihoods on the Bekenstein-Hod bound, $P(\mathrm{D}_i \, | \, \mathcal{H}_i < 1)$, corresponding to the case where a uniform prior is set on the $\mathcal{H}$ parameter. As expected, the two estimates show a closer agreement on loud events, while for quieter ones they differ. For the latter, the likelihoods are only mildly informative and thus extend to large values of the $\mathcal{H}$ parameters, hence yielding weak bounds. In summary, no significant disagreement with the theoretical prediction is found.  

\begin{table}
\caption{Probabilities that the Bekenstein-Hod bound holds in BBHs observed by LIGO and Virgo, extracted from the posterior ($P_{\mathcal{H} | \mathrm{D}}$) and likelihood ($P_{\mathrm{D} | \mathcal{H}}$) distributions. The combined estimates are computed, respectively, from the median posterior distribution of a hierarchical analysis (see main text) and the evidence-weighted likelihood average of the considered events. A high agreement with the prediction of GR, BH thermodynamics and information theory is found.}
\begin{ruledtabular}
\begin{tabular}{lll}
Event & $P_{\mathcal{H}|\mathrm{D}}$ & $P_{\mathrm{D} | \mathcal{H}}$ \\
\hline
GW150914           & 0.91 & 0.75 \\
GW170104           & 0.70 & 0.28 \\
GW190513\_205428   & 0.74 & 0.24 \\
GW190519\_153544   & 0.72 & 0.43 \\
GW190521           & 0.99 & 0.98 \\
GW190521\_074359   & 0.97 & 0.86 \\
GW190602\_175927   & 0.74 & 0.44 \\
GW190910\_112807   & 0.99 & 0.94 \\
Combined 		  & $0.94^{+0.05}_{-0.14}$ & 0.93 \\
\end{tabular}
\label{tab:probs}
\end{ruledtabular}
\end{table}

We extract a combined statement from all the available events by computing the probability that the \textit{whole population} of analysed massive remnant BHs respects the bound. We do so by means of a \textit{hierarchical modeling}~\cite{gelman2013, Isi:2019asy} procedure.
Each remnant BH possesses a value of $\mathcal{H}$ dependent on the GW signal source parameters, yielding a different posterior distribution, denoted by $\{ p(\mathcal{H}_i |\mathrm{D}_i) \}_{\scriptscriptstyle i \in [1,N] }$, where $N$ is the total number of observed events. However, if GR, information theory, and BH thermodynamics are valid, all the $p(\mathcal{H}_i |\mathrm{D}_i)$ need to have negligible support besides unity.
To verify this prediction, we assume that the $\mathcal{H}_i$ are realisations of an underlying \textit{parent} distribution $p(\mathcal{H} \, |\, \mathrm{D}^N)$, where $\mathrm{D}^N := \{ \mathrm{D}_1, \dots, \mathrm{D}_N \} $. This observed distribution $p(\mathcal{H} \, |\, \mathrm{D}^N)$, will depend on the BH dynamics, and on the properties of the observed BHs masses and spins.
The actual functional form of the distribution is not too relevant, as long as it is: i) sensitive to violations of the bound; ii) flexible enough to incorporate the structure of the observed distributions. To meet both these conditions, we assume the uni-variate version of a Dirichlet distribution, the Beta distribution $p(\mathcal{H} \, |\, \mathrm{D}^N) = \beta(\widetilde{\mathcal{H}} \,; a,b)$, where $a,b>0$ are the concentration parameters. Since the support of the beta distribution is restricted to the unit interval, we renormalised the $\mathcal{H}$-support to the largest value found in the entire set of events considered, $\widetilde{\mathcal{H}} := \mathcal{H}/\mathcal{H}^{max}$. The Beta distribution is a convenient way of expressing the underlying parent distribution $p(\mathcal{H} \, |\, \mathrm{D}^N)$ in a parametric form. By varying the concentration parameters of the distribution, its shape can model both situations where there is uniform support over the whole $\mathcal{H}$ range and situations where the support of $\mathcal{H}$ is peaked below or above the Bekenstein-Hod bound. We verified the robustness of our conclusions by repeating the analysis presented below using alternative parametrizations, such as a Gamma distribution. None of our conclusions are affected. 
The hierarchical approach also provides a robust and convenient framework to eliminate the prior dependence on our final measurement. This is especially important since, as already discussed, the prior on the $\mathcal{H}$ parameter obtained by imposing the standard priors assumed in LVC analyses, peaks below one, thus it favors a-priori GR and can mask possible deviations in the BH information emission rate for quiet events.
The inference task then reduces to the problem of inferring the values of the concentration parameters $a,b$ from the available data.

To this end, we use Bayes theorem and the quantitative rules of bayesian inference~\cite{Jaynes2003}. We start from the global posterior distribution of all the nuisance ($\mathcal{H}_i$) and hyper-parameters ($a,b$):  
\begin{equation}\label{eqn:posterior}
p(\mathcal{H}^N, a, b | \mathrm{D}^N) = \frac{p(\mathcal{H}^N, a, b) \, p(\mathrm{D}^N | \mathcal{H}^N, a, b) }{Z} \text{,}
\end{equation}
where $\mathcal{H}^N := \{ \mathcal{H}_i \}_{\scriptscriptstyle i \in [1,N] }$ and $Z$ is the evidence normalisation factor.
The prior splits in two terms:
\begin{equation}\label{eqn:prior}
p(\mathcal{H}^N, a, b) = p(\mathcal{H}^N | a, b) \, p(a, b) \text{,}
\end{equation}
where $p(a,b)$ is chosen to be a uniform distribution. The first factor entering Eq.~\eqref{eqn:prior} encodes our hierarchical model assumption that each $\mathcal{H}_i$ originates from a Beta distribution:
\begin{equation}
p(\mathcal{H}^N | a,b) = \prod_i p(\mathcal{H}_i | a,b) = \prod_i \beta(\widetilde{\mathcal{H}}_i \,; a,b) \text{,}
\end{equation}
where $\beta(x \,; a,b)$ is the Beta function and we made use of the independence of different GW events.

\begin{figure}[t]
\includegraphics[width=0.5\textwidth]{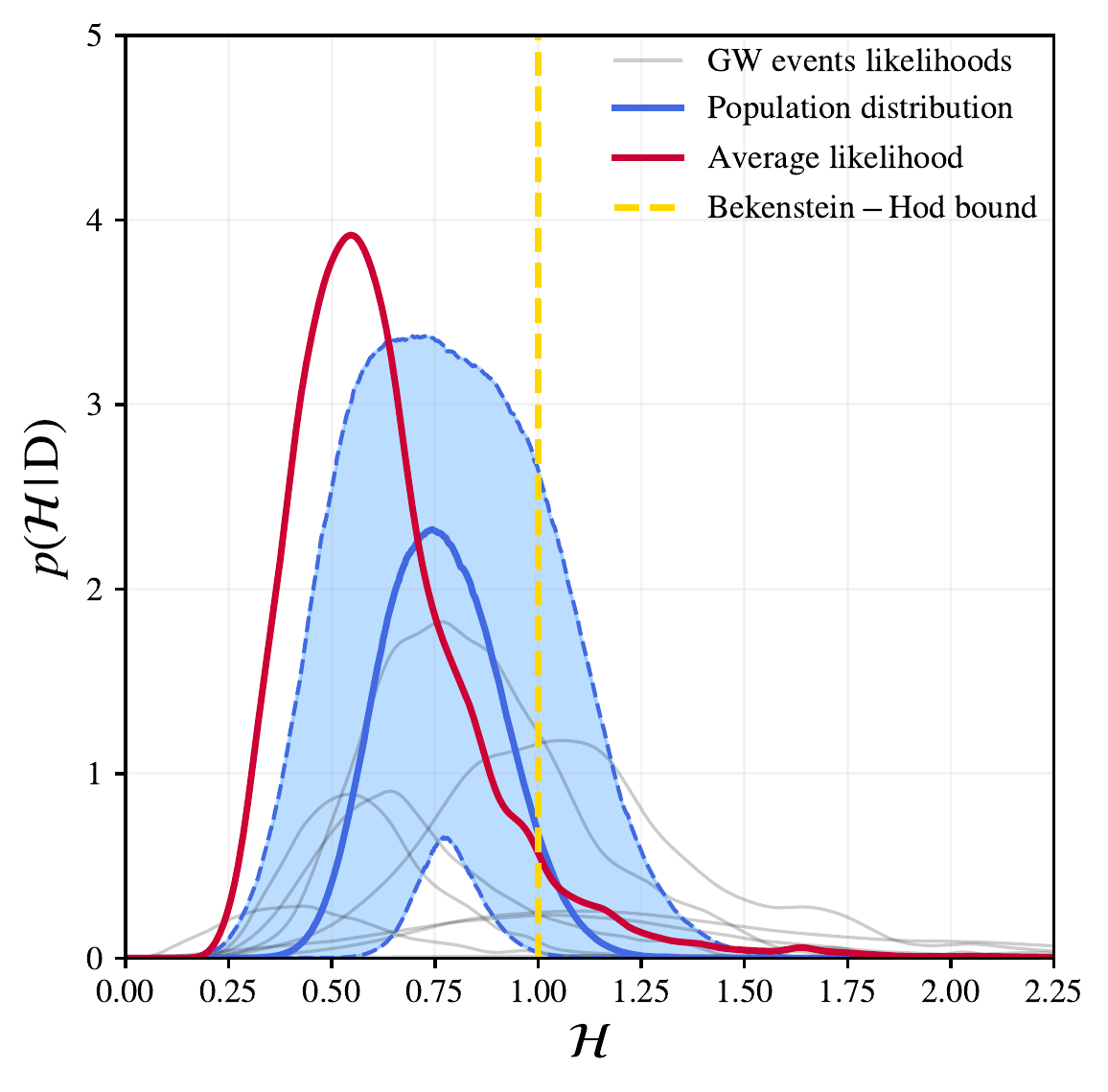}
\caption{Median and $90\%$ credible levels on the Bekenstein-Hod parameter $\mathcal{H}$ parent distribution, obtained through a hierarchical model (blue area). Single-events likelihood (grey curves) are also displayed, together with their evidence-weighted average (red curve). The probability that the bound (gold dashed line) is obeyed by the whole population are $\tilde{p} = 0.94^{+0.05}_{-0.14}$ when assuming the posterior distribution and $\tilde{p} =  0.93$ when assuming the average likelihood.}
\label{fig:bound2} 
\end{figure}

The likelihood can be further simplified using the fact that the events are independent and that the hyper-parameters do not enter directly in the single-event likelihoods:
\begin{equation}
p(\mathrm{D}^N | \mathcal{H}^N, a, b) = p(\mathrm{D}^N | \mathcal{H}^N) = \prod_i p(\mathrm{D}_i | \mathcal{H}_i) \text{.}
\end{equation}
Each single-event likelihood $p(\mathrm{D}_i | \mathcal{H}_i)$ is constructed from a Gaussian Kernel Density Estimation~\cite{scipy} applied on the posterior and prior distributions of each $\mathcal{H}_i$.
We use the CPNest Nested Sampling algorithm~\cite{CPNest} to extract the posterior probability distribution of Eq.~\ref{eqn:posterior}. An additional complication arises from the fact that the $a,b$ parameters are strongly correlated, giving origin to upper prior railing in the $b$ parameter, independently of the chosen upper prior bound $b_{max}$. 
We thus choose to marginalise over the values of $a_{max}, b_{max}$ by repeating the analysis for different choices of the parameters ($a_{max}=b_{max} = \{1,5,10,50,100,500,1000\})$ and selecting the result yielding the largest evidence. 
Our results are not affected by the observation of this railing. As the upper prior bounds increase, the posterior tends to shrink towards values more and more in agreement with the Bekenstein-Hod bound. Consequently, by selecting the highest evidence case ($a_{max}=b_{max}=100$), we employ a conservative choice. In Fig.~(\ref{fig:bound2}) we display, in blue, the median and $90\%$ credible intervals of the posterior probability distributions $p(\mathcal{H}| \mathrm{D}^N, a, b)$.
Single events likelihood are shown in grey, for comparison.
We finally compute the probability that the Bekenstein-Hod bound (gold vertical line) is obeyed on a population level, by computing the $p$-value ($\tilde{p} := p(\mathcal{H} < 1 | \mathrm{D}^N, a, b)$) for each of $a,b$ sample obtained from Eq.~\eqref{eqn:posterior}. The result yields a $\tilde{p}$-value distribution strongly peaked towards unity with median and $90\%$ credible levels given by $\tilde{p} = 0.94^{+0.05}_{-0.14}$, where $\tilde{p} = 1$ would indicate perfect agreement with the prediction, while $\tilde{p} = 0$ perfect disagreement. The Bekenstein-Hod bound is respected with very high confidence by the observed BBH population. 
As an additional check, we compared our result with the corresponding value coming from a naive point-estimate of the average $\mathcal{H}$ likelihood, the latter being insensitive to specific hierarchical modeling choices. A weighted average over single events likelihoods, with weights given by the respective evidences, yields the red curve displayed in Fig.~(\ref{fig:bound2}), corresponding to $\tilde{p} =  0.93$. The excellent agreement between this un-modelled estimate and the median of the hierarchical population posterior confirms the robustness of the adopted population model.

\textit{Conclusions--} BHs are expected to be the fastest-dissipating objects in the Universe, 
in the sense that they possess the shortest possible relaxation time for a given temperature~\cite{Hod:2006jw}.  
In this Letter, we obtained an observational verification of the \textit{Bekenstein-Hod 
information emission bound} using a Bayesian time-domain analysis applied to the binary black holes of the LIGO-Virgo GWTC-2 catalog. 
The result is consistent with the predictions of GR, BH thermodynamics and information theory. Our analysis
provides the first experimental verification of a long standing prediction on the dynamical information-emission process of a BH.\\

\begin{acknowledgments}

\textit{Software} Open-software \texttt{python} packages, accessible through \texttt{PyPi}, used in this work comprises: \texttt{corner, gwsurrogate, h5py, matplotlib, numba, numpy, scipy, seaborn, surfinBH} \cite{corner, gwsurrogate, hdf5, matplotlib, numba, numpy, seaborn, surfinbh1, surfinbh2}.\\

\textit{Acknowledgments}
The authors would like to thank Aditya Vijaykumar, Giancarlo Cella, and Tjonnie Li for stimulating discussions and Huan Yang, Alessandro Pesci, Roman Konoplya for comments on the manuscript.
J.V. was partially supported by STFC grant ST/K005014/2.
This research has made use of data, software and/or web tools obtained from the Gravitational Wave Open Science Center (https://www.gw-openscience.org), a service of LIGO Laboratory, the LIGO Scientific Collaboration and the Virgo Collaboration. LIGO is funded by the U.S. National Science Foundation. Virgo is funded by the French Centre National de Recherche Scientifique (CNRS), the Italian Istituto Nazionale di Fisica Nucleare (INFN) and the Dutch Nikhef, with contributions by Polish and Hungarian institutes.
\end{acknowledgments}

\bibliographystyle{apsrev}
\bibliography{Hod_bound_Bibliography}

\end{document}